\documentclass[12pt,preprint]{aastex}

\def\ni{\noindent}
\def\s{{\rm s}}

\def\cm{{\rm\,cm}}

\def\gm{{\rm\,g}}
\def\g{{\rm\,g}}

\def\rad{{\rm rad}}

\def\pomega{\tilde{\omega}}

\begin{document}

\shortauthors{Chiang and Culter}
\shorttitle{Narrow Ring Dynamics}

\title{Three-Dimensional Dynamics of Narrow Planetary Rings}

\author{Eugene I. Chiang \& Christopher J. Culter}

\affil{Center for Integrative Planetary Sciences\\
Astronomy Department\\
University of California at Berkeley\\
Berkeley, CA~94720, USA}

\email{echiang@astron.berkeley.edu, cculter@uclink.berkeley.edu}

\begin{abstract}
Narrow planetary rings are eccentric and inclined.
Particles within a given ring must therefore share the same
pericenter and node. We solve for the three-dimensional geometries
and mass distributions that enable the Uranian $\alpha$
and $\beta$ rings, and the Saturnian Maxwell and Colombo
(Titan) rings, to maintain simultaneous apsidal and nodal lock. Ring
self-gravity, interparticle collisions, and the quadrupole
field of the host planet balance each other to achieve
this equilibrium. We prove that such an equilibrium
is linearly stable. Predictions for the Saturnian ringlets
to be tested by the Cassini spacecraft include: (1) ringlet
masses are of order a few $\times \, 10^{19} \gm$,
(2) surface mass densities should increase from ring
midline to ring edges, and (3) rings are vertically warped
such that the fractional variation of inclination across
the ring is of order 10\%. Analogous predictions are made
for the Uranian rings. Simultaneous apsidal and nodal locking
forces the narrowest portion of the ring---its ``pinch,'' where
self-gravitational and collisional forces are strongest---to
circulate relative to the node,
and introduces previously unrecognized time-varying forces perpendicular
to the planet's equator plane.
We speculate that such periodic stressing might
drive kilometer-scale bending waves at a frequency
twice that of apsidal precession; such flexing might be
observed over a few weeks by Cassini.
\end{abstract}


\section{INTRODUCTION}
\label{intro}
The $\alpha$ and $\beta$ rings of Uranus are eccentric and inclined (Elliot et
al.~1984; French et al.~1991).
The mean eccentricity, $\bar{e}$, of each ring is
$(0.761 \pm 0.004) \times 10^{-3}$
and $(0.442 \pm 0.003) \times 10^{-3}$,
respectively. The mean inclination, $\bar{I}$,
of each ring with respect to the equator plane of Uranus
is $0.0152 \pm 0.0006$ deg and 0.0051 $\pm 0.0006$ deg, respectively.
To maintain an observed eccentricity and inclination, each ring
must be composed of particles that share the same longitude of pericenter
and same longitude of ascending node. The alignments of apsides and of nodes
must simultaneously follow from a delicate balance of forces due to
the gravitational field of the planet, ring self-gravity, and
interparticle collisions. Obtaining a complete equilibrium solution
for a ring's three-dimensional geometry and mass distribution
has been a goal of dynamicists for decades. The solution
bears directly on ring ages and origins (see, e.g.,
Goldreich \& Tremaine 1979a; Chiang 2003).

The Maxwell and Colombo ringlets are
Saturnian analogs of the Uranian $\alpha$ and $\beta$ rings (Porco
1990).\footnote{The Colombo ringlet is also known informally as the Titan
ringlet.}
The mean eccentricities of the Saturnian ringlets,
but not their mean inclinations, are measured from Voyager
spacecraft observations. Table \ref{param} summarizes observed ring
parameters.

Shepherd satellites are expected to confine narrow rings and maintain
the latter's sharp edges (Borderies, Goldreich, \& Tremaine 1982), but
no satellite has yet been observed to accompany any of the
aforementioned ringlets. This state of affairs promises to
change with the arrival of the Cassini spacecraft to Saturn in 2004.
Narrow rings and their attendant shepherds
furnish the most accessible laboratories
we have for studying disk-satellite interactions. Their study
informs problems on grander scales,
including migration of planets within circumstellar disks (see, e.g.,
Chiang 2003).

The steady-state condition of apsidal alignment, combined with
measurements of the ring eccentricity profile, $e(a)$, where $a$ is
the semi-major axis of a ring streamline,
can be employed to calculate the ring surface
density, $\Sigma (a)$. First attempts at this calculation account
for the planetary quadrupole field and ring self-gravity,
but omit the effects of interparticle collisions
(Goldreich \& Tremaine 1979b, hereafter, the ``standard
self-gravity model''). Chiang \& Goldreich (2000, hereafter CG00)
restore the latter effects to demonstrate how collisions near ring
boundaries can dramatically raise the ring masses required for
apsidal alignment. Surface densities derived by accounting for pressure
forces exceed those derived without pressure forces
by 1--2 orders of magnitude;
the larger masses can be reconciled with occultation measurements
and with classical theories of ring shepherding (CG00).
A more careful treatment of pressure forces in the ring interior
is undertaken by Mosqueira \& Estrada (2002), who reach the same
qualitative conclusions as those of CG00.

All the works cited above consider only
apsidal alignment and neglect nodal alignment.
Nodal alignment, but not apsidal alignment,
is studied by Borderies, Goldreich, \& Tremaine (1983a),
who consider ring self-gravity and planetary gravity, but neglect
collisions. Ring eccentricities are set to zero
in their analysis.

The ring's true surface density profile must be simultaneously
reconciliable with both the alignment of apsides and the alignment of nodes;
the horizontal structure of a narrow ring is entwined with its
vertical structure. This paper seeks
to simultaneously treat apsidal and nodal alignment
while accounting for the full panoply
of forces due to the planetary quadrupole field,
ring self-gravity, and interparticle collisions.
In \S\ref{equilibrium}, we derive equilibrium
ring surface densities and vertical geometries that lock the
apsides and nodes of a given ring. We apply our solutions
to the $\alpha$ and $\beta$ rings of Uranus, and the Maxwell and Colombo
ringlets of Saturn. In \S\ref{stability}, we
present a proof that circular, nodally locked rings are linearly stable
to perturbations to their inclinations and nodes. The beginnings
of such a proof can be found in Borderies, Goldreich, \& Tremaine (1983b);
here, we state the arguments more completely and explicitly.
In \S\ref{discussion}, we discuss our results, highlighting
the future impact of the Cassini spacecraft on studies of narrow rings
and unresolved theoretical issues.

\section{EQUILIBRIUM}
\label{equilibrium}

Our procedure for deriving the mass and 3-dimensional structure of a narrow
ring is summarized as follows. The range of semi-major axes spanned by the
ring, the eccentricity profile [$e(a)$], and the mean inclination ($\bar{I}$)
are assumed to be given. From $e(a)$,
we compute the surface density profile, $\Sigma(a)$, by enforcing
apsidal alignment across the ring and
by accounting for planetary oblateness, ring
self-gravity, and interparticle collisions.\footnote{The Colombo
ringlet has the added complication that it inhabits a 1:0 apsidal
resonance established by the Saturnian satellite, Titan; the ring's apsidal
precession frequency matches the mean motion of Titan. It is straightforward
to show that the contribution of Titan to the differential apsidal
precession rate across the Colombo ringlet is small compared to the
differential apsidal precession
induced by the quadrupole field of Saturn.
We will therefore neglect perturbations by
Titan on the Colombo ringlet in this paper.} 
This computation is
described in detail in \S\ref{surf}. Next, from $\Sigma(a)$
and $\bar{I}$, we compute the inclination profile, $I(a)$,
by enforcing nodal alignment across the ring and by accounting
for planetary oblateness and ring self-gravity but not
interparticle collisions. The computation of $I(a)$
is described in \S\ref{inca}. Finally, in \S\ref{check},
we gauge {\it a posteriori}
the extent to which our neglect of collisional
stresses in the derivation of $I(a)$ is justified.

Our computational procedure assumes the ring
is characterized
by a time-independent argument of pericenter, $\omega$.
This means the narrowest portion of the ring (its pericenter)
lies at a fixed angle relative to the ring's node on the equator
plane of the planet.
Strictly speaking, the assumption of stationary $\omega$ is invalid,
since

\begin{equation}
\dot{\omega} = \dot{\pomega} - \dot{\Omega} = 2 \dot{\pomega} \neq 0 \, .
\end{equation}

\ni The common precession rate of the longitude of pericenter of all apsidally
aligned streamlines,
$\dot{\pomega}$, is set by the quadrupole field of the central
planet and is positive (in the direction of increasing true anomaly).
The common precession rate of the longitude
of ascending node of all nodally aligned streamlines, $\dot{\Omega}$,
is also set by the planetary quadrupole field,
is identical in magnitude to $\dot{\pomega}$,
and is negative.
That $\dot{\omega} \neq 0$ implies that the pericenter
of the ring circulates relative to the node on the planet's
equator plane; since self-gravitational forces are largest at pericenter,
the contribution to differential nodal precession
from ring self-gravity cannot be time-independent.
Our hope is that this complication does not introduce serious errors
into our calculation of the time-averaged ring geometry; 
we point out below where our assumption of fixed argument
of pericenter is employed, explore in a preliminary
but quantitative manner the consequences of its violation 
in the Appendix, and speculate about its true effects
in the discussion section (\S\ref{discussion}).

\subsection{Surface Density}
\label{surf}

We divide the ring into a set of $2N$ wires and compute the mass of each
wire by imposing the steady-state
condition of apsidal alignment. The $2N$ wires are equally spaced
in semi-major axis; $N$ wires
lie at semi-major axes smaller than that of the ring midline,
and $N$ wires lie at greater semi-major axes. Our calculation
differs from that of CG00 only in the treatment of collisional forces.
We restrict ourselves here to describing this difference; for background
details, see CG00.

CG00 highlight the importance of interparticle collisions at
ring boundaries and account crudely for their effects by introducing
force terms in the equations of motion for ``endwires'' located within
a collisional mean free path, $\lambda$,
of either ring boundary. No account is made
of collisional stresses in the ring interior (regions displaced
many mean free paths from either boundary) in the quantitative
model they present. Here we improve upon their calculation by making
such an account, although still in a crude and prescriptive manner.
We take the solution of CG00 for $\Sigma(a)$ and derive from it
a collisional acceleration everywhere within the ring:

\begin{equation}
\mathbf{C}(a,f) = -\frac{1}{\Sigma} \frac{\partial (\Sigma c^2)}{\partial a}
\frac{S}{1 - q_e \cos f} \, \mathbf{\hat{r}} \, ,
\label{collacc}
\end{equation}

\ni where $\mathbf{\hat{r}}$ points in the radial direction, $q_e = a de/da$
is
the dimensionless eccentricity gradient, and $f$ measures true anomaly.
The factor $1 - q_e \cos f$ is proportional to the local radial spacing
between wires (see, e.g., Goldreich \& Tremaine 1979b). Following
CG00, we adopt a constant $q_e(a) = \bar{q_e}$ across
the entire ring. The velocity
dispersion, $c$, and a dimensionless factor, $S$, are described as follows.

We prescribe the following profile for the square of the velocity dispersion,

\begin{equation}
\label{collprescrip}
c^2(a) = c_i^2 + c_b^2 \exp{(-|a-a_b|/w_r)} \, ,
\end{equation}

\ni where $c_i$, $c_b$, and $w_r > 0$ are the same constants as employed
by CG00, and $a_b$ is the semi-major axis of the ring boundary that
is nearest a given wire at $a$. According to (\ref{collprescrip}),
the velocity dispersion
decreases from either ring boundary towards the ring interior
over a lengthscale $w_r$.
The physics underlying the enhancement of velocity
dispersion near ring boundaries is elucidated by
Borderies, Goldreich, \& Tremaine
(1982) and estimates for $c_i$, $c_b$, and $w_r$ are derived by CG00.
The actual values we employ are contained in Table \ref{param2}.

The function

\begin{equation}
S = \exp{[-2\lambda/( |a-a_b| + \lambda) ]}
\end{equation}

\ni is a softening parameter that we impose because the true acceleration
due to particle collisions near a ring boundary is likely
overestimated by the usual hydrodynamic expression for the acceleration due to
pressure gradients [$-\Sigma^{-1} \nabla (\Sigma c^2)$].
Within a few mean free paths, $\sim$$\lambda$, of the ring edge,
the hydrodynamic approximation breaks down and particles behave
more ballistically with less regard for large-scale gradients in the surface
density. The parameter $S$ quickly grows from $e^{-2}$ to unity as we recede
from the ring edge towards the ring midline.

The collisional acceleration, $\mathbf{C}$, is inserted into Gauss's equation
for a given ring particle's apsidal precession rate, $d\pomega/dt$,
and averaged over true anomaly to yield, for a given wire,
$\langle d\pomega/dt \rangle_C$.
This collisional contribution to the wire's precession rate
adds to other contributions due to planetary oblateness
and inter-wire gravity. Expressions for the latter two contributions
are supplied by CG00. The condition of
apsidal alignment requires that precession rates
of all wires be equal to the precession rate of a test particle
at the ring midline, and yields $2N$ linear equations for the
$2N$ wire masses. The solution of this linear system generates a new surface
density profile, $\Sigma'(a)$. The entire calculation is then
repeated: from this new surface density,
we calculate a new collisional acceleration profile
[$\mathbf{C'}(a,f)$], a new set of wire precession rates
due to pressure forces [$\langle d\pomega/dt\rangle_{C'}$],
and a new set of $2N$ linear equations.
In this way, the solution for the surface density
is iterated until convergence is achieved.

In practice, for model parameters appropriate to the Uranian $\alpha$
and $\beta$ rings and the Saturnian Maxwell and Colombo ringlets
(see Table \ref{param2}), we find that the solution of
$2N = 2000$--6000 wire masses converges after $\sim$10 iterations.
Because the solution is reflection-symmetric about the ring midline,
only $N = 1000$--3000 linearly independent equations need be solved
for $N$ distinct wire masses.
The derivative in equation (\ref{collacc})
is computed numerically using a Savitzky-Golay smoothing filter
of order 2 and having a width of 20 wires (Press et al.~1992).
The system of $2N$ linear equations
is solved using subroutine DGESV of the LAPACK (Linear Algebra
Package) software library. Because many of our parameters
such as $c$ and $S$ are only order-of-magnitude estimates,
our profiles are probably accurate to factors of a few, at best.
Nevertheless, by accounting not only for interparticle collisions
near ring boundaries but also for collisions within the ring interior,
we may explore the qualitative effects of incorporating the latter
and make a first-cut correction to the solutions obtained by CG00.

Computed surface density profiles for the $\alpha$, $\beta$, Maxwell,
and Colombo ringlets are displayed in Figures 1--4, respectively.
Near the ring midline, our solutions require less mass than those
of CG00, a consequence of local pressure gradients that compress
the ring and abet ring self-gravity. By contrast, our computed
peaks in surface density near ring boundaries are larger than those
of CG00. The reason for this is as follows. As a given peak in surface
density is approached from the ring midline, steep
and inward-directed pressure forces must be balanced by
the outward gravitational attraction of massive endwires on the far side
of the peak.
These inward-directed forces are neglected by CG00; accounting for
them here
leads to endwire masses larger than those obtained by CG00.

These same qualitative conclusions are reached by Mosqueira \& Estrada (2002),
who employ various prescriptions for collisional stresses
that differ from ours. Their surface density profiles
deviate from ours by large amounts, sometimes by more than an order
of magnitude, reflecting the sensitivity of the shape of the profile
to choice of boundary conditions. Some consolation
may be had in the finding of Mosqueira \& Estrada (2002)
that the total ring mass, $M$, is less sensitive to this choice.
The chief shortcoming of all calculations of the surface density
profile, including our own,
is the prescriptive and slightly
arbitrary description of the ring boundary. We have assumed here
that $q_e$ is constant across the entire ring, even in boundary regions
resonantly perturbed by shepherd satellites where streamlines
can no longer be described as simple ellipses. See CG00 for a discussion
of this point.

\placefigure{fig1}
\begin{figure}
\epsscale{0.9}
\plotone{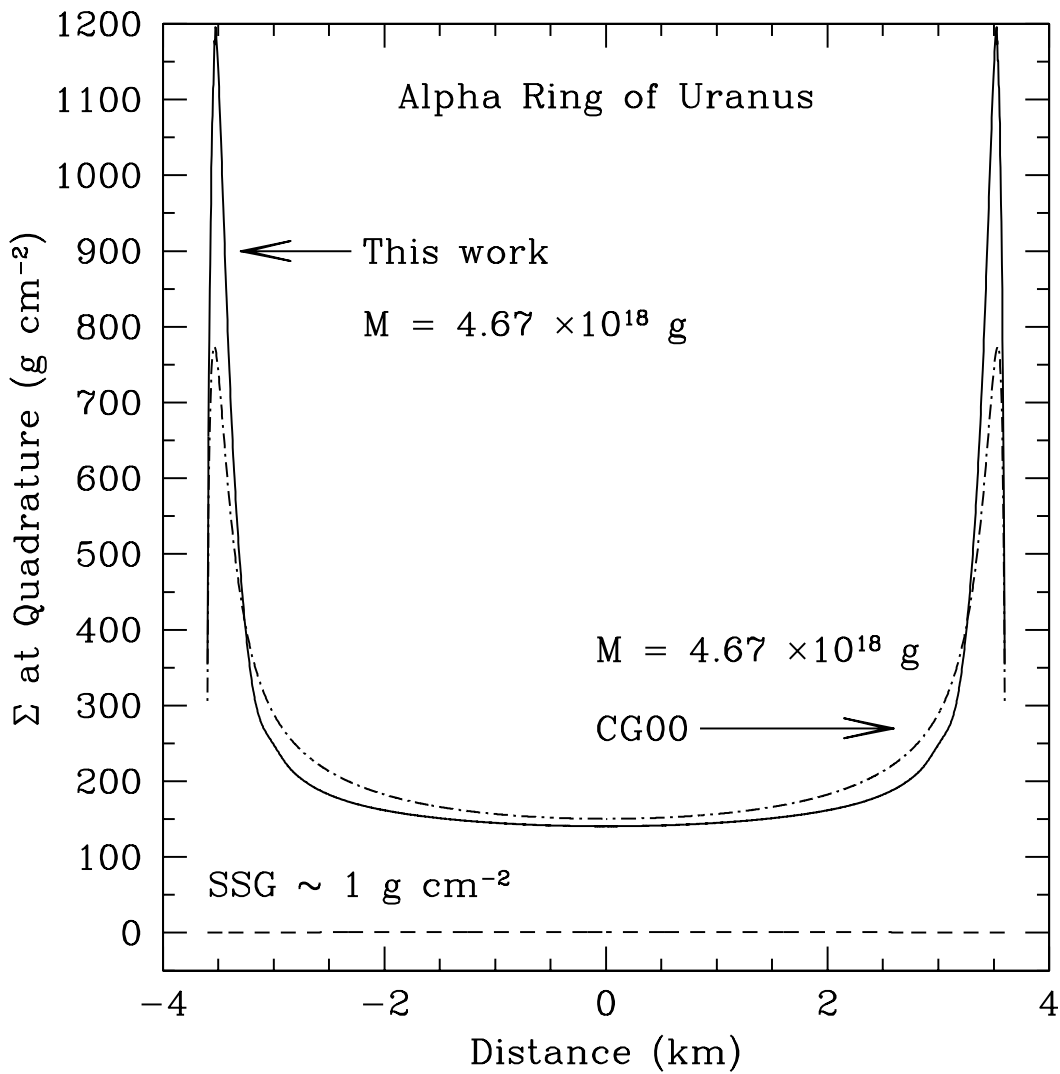}
\caption{Computed surface density profiles for the $\alpha$ ring of
Uranus, for $q_e(a) = \bar{q_e} = 0.472$.
Including pressure forces in the ring interior lowers
the surface density near the ring midline and raises it near the ring
boundary, as compared to the model by CG00 who neglect
interior pressure gradients. The total mass of the ring, $M$, hardly changes
between models, however. The bottom dashed line corresponds
to the standard self-gravity (SSG) model of Goldreich \& Tremaine (1979b);
it predicts surface densities too low to accord with observations (CG00).
\label{alpha}}
\end{figure}

\placefigure{fig2}
\begin{figure}
\epsscale{0.9}
\plotone{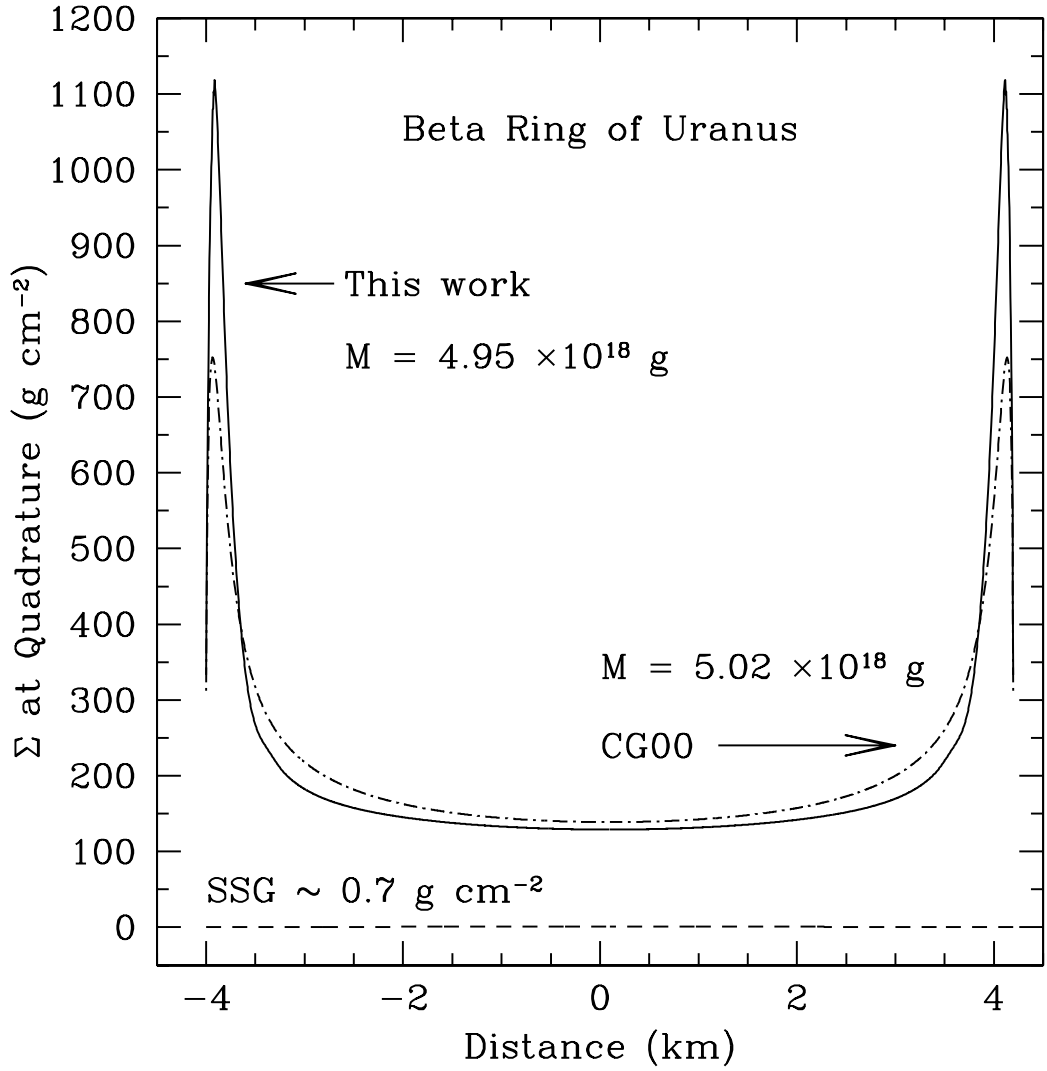}
\caption{Same as Figure \ref{alpha}, but for the $\beta$ ring of Uranus
($q_e = \bar{q}_e = 0.370$).
\label{beta}}
\end{figure}

\placefigure{fig3}
\begin{figure}
\epsscale{0.9}
\plotone{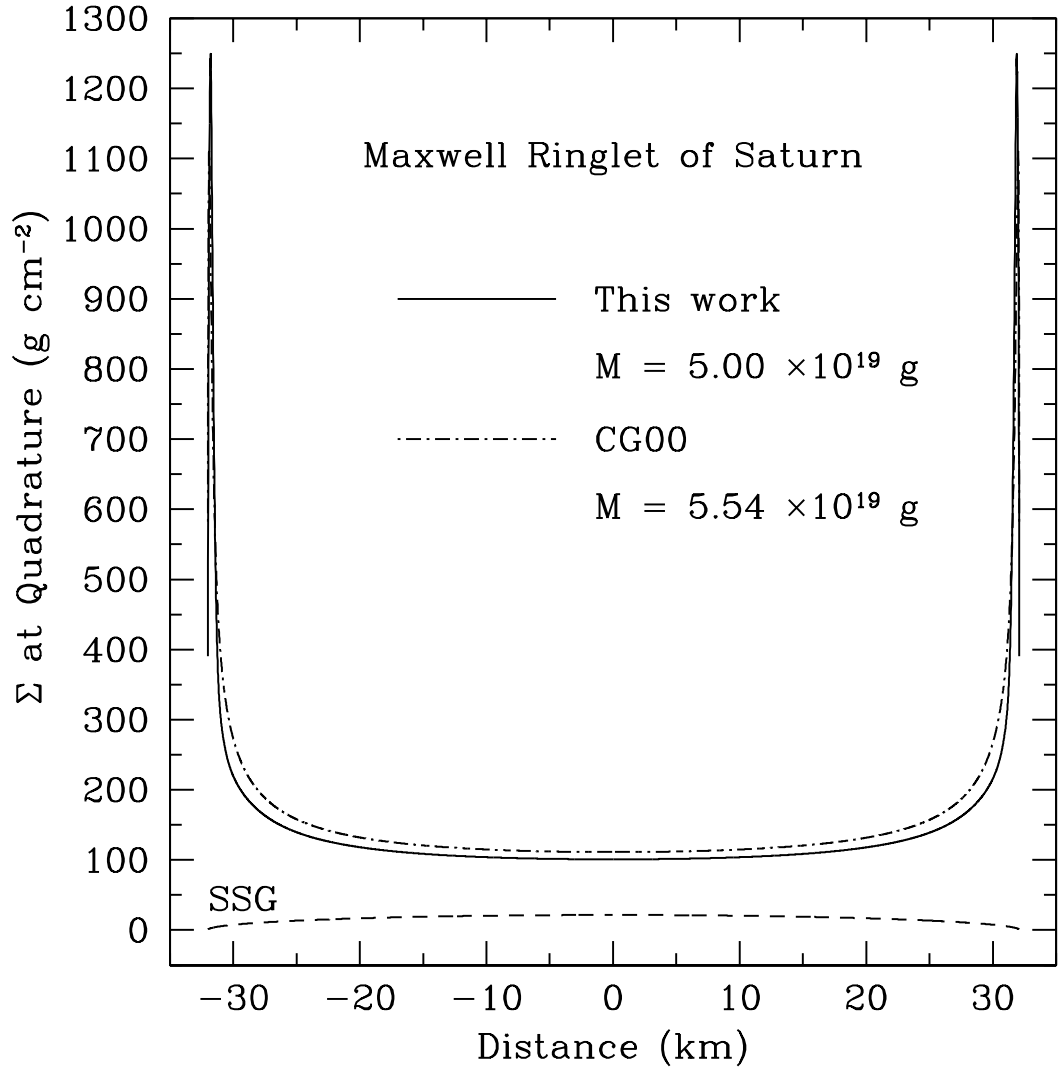}
\caption{Same as Figure \ref{alpha}, but for the Maxwell ringlet of Saturn
($q_e = \bar{q}_e = 0.46$).
\label{maxwell}}
\end{figure}

\placefigure{fig4}
\begin{figure}
\epsscale{0.9}
\plotone{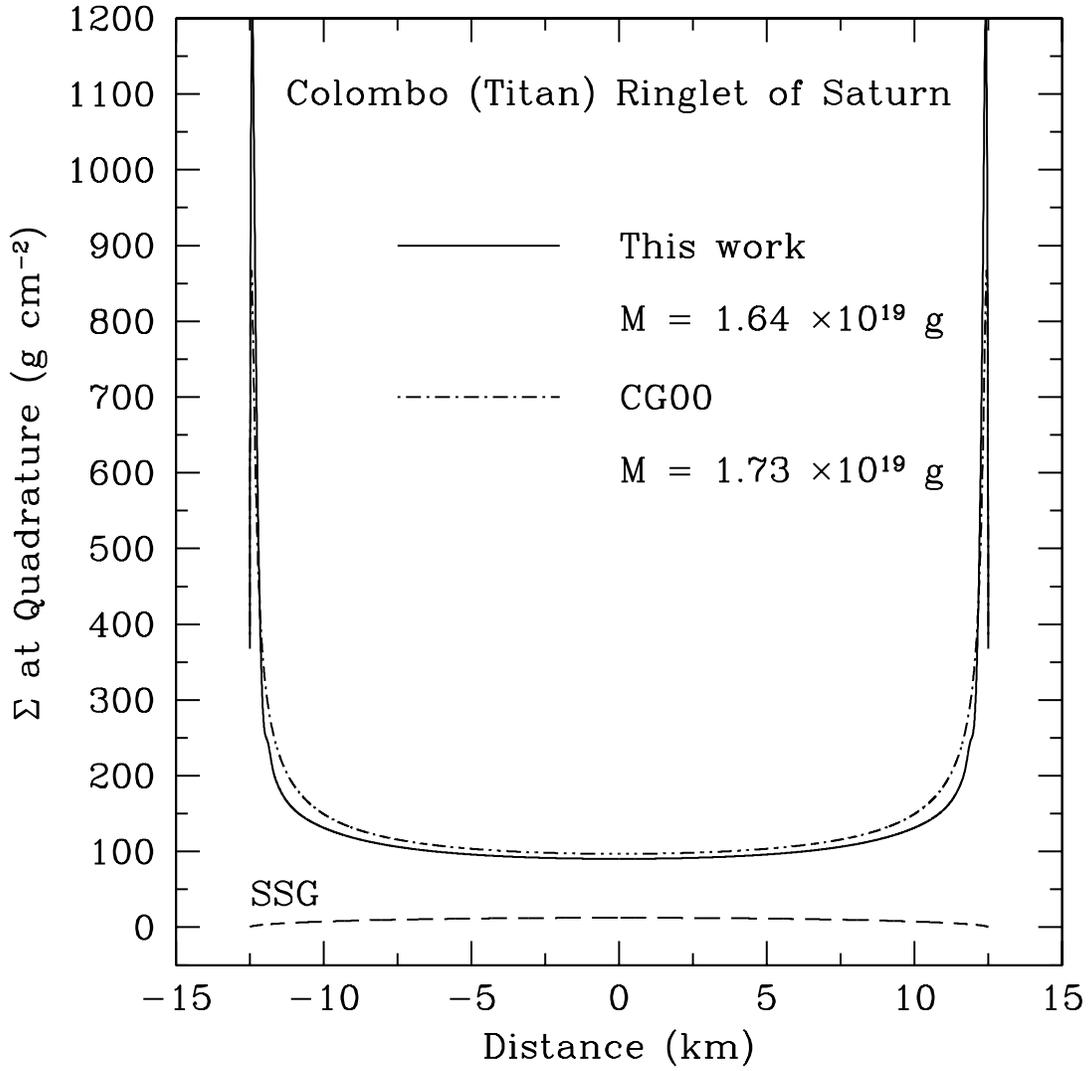}
\caption{Same as Figure \ref{alpha}, but for the Colombo (Titan) ringlet of
Saturn ($q_e = \bar{q}_e = 0.44$).
\label{colombo}}
\end{figure}

\subsection{Inclination Profile}
\label{inca}
Given $\bar{I}$ and $\Sigma(a)$, we compute $I(a)$
by imposing the steady-state condition of nodal alignment.
We now discretize the ring into $2N-1$ wires that share
the same longitude of ascending node and solve
for the orbital inclination of each wire with respect
to the equator plane of the planet.
Note that
whereas the calculation of the surface density in \S\ref{surf}
employs an even number of wires ($2N$), here we employ
an odd number for the calculation
of the inclination profile.
The reason for this change of procedure
is to have the $N^{\rm{th}}$ wire lie on the ring midline
and to assign $I_N = \bar{I}$. With this
reference point defined, the inclinations of the
$N-1$ wires that lie to one side of the midwire
may then be computed. Inclinations of the
$N-1$ wires that lie on the other side of the midwire
follow from reflection anti-symmetry about the midwire.

The nodal precession rate of the $j^{\rm{th}}$ wire
relative to that of a particle on the ring midline
and due to the quadrupole field of the central planet is given by

\begin{equation}
\Delta_Q \langle \dot{\Omega} \rangle _j = \frac{21}{4} J_2 \bar{n} \left(
\frac{R_p}{\bar{a}} \right)^2 \frac{a_j - \bar{a}}{\bar{a}} \, ,
\end{equation}

\ni where $\bar{a}$ and $\bar{n}$ are the semi-major
axis and mean motion, respectively, of a particle on
the ring midline, $J_2$ is the usual dimensionless measure
of the planetary quadrupole field, $R_p$ is the planetary radius,
and $\langle \rangle$ denotes a time-average over one orbit.
Here $j$ runs from 1 (the wire closest to the planet) to
$2N-1$ (the wire furthest from the planet).

The corresponding differential nodal precession rate due to inter-wire
gravity is

\begin{equation}
\Delta_G \langle \dot{\Omega} \rangle_j = \frac{\bar{n}\bar{a}}{\pi M_P
\bar{I}} \sum_{k\neq j}^{2N-1} \frac{m_k}{a_k-a_j} B q_{kj} \, ,
\label{igrav}
\end{equation}

\ni where


\begin{equation}
q_{kj} = q_{jk} = \bar{a} \frac{I_k - I_j}{a_k - a_j} \, ,
\end{equation}

\ni $M_P$ is the mass
of the planet, $m_k$ is the mass of the $k^{\rm{th}}$ wire, 
and $B$ is a dimensionless function of order unity that depends on the
eccentricity profile of the ring, the inclination
profile ($q_{jk}$), and the arguments of pericenters of the various wires.
Following our assumption of a time-independent $\omega$ (see the discussion
preceding \S\ref{surf}), we will take $B = 0.77$ constant.
Though $B$ actually varies
with time because $\dot{\omega} \neq 0$, its maximum range of variation
is less than a factor of 2, as we derive in the Appendix.

The $N$ distinct wire masses computed in \S\ref{surf} furnish
the set of $\{ m_j \}$ used in equation (\ref{igrav})
from $j = 1$ to $N$; the remaining $N-1$ wire
masses are assigned by reflection symmetry across the ring midline.

Steady nodal alignment implies

\begin{equation}
\Delta_Q \langle \dot{\Omega} \rangle_j + \Delta_G \langle \dot{\Omega}
\rangle_j = 0 \,\,\,\,\,\,\,\,\,\, \{j = 1, \ldots, N-1\} \, ,
\label{masterinc}
\end{equation}

\ni where we have neglected the contribution from interparticle
collisions; the validity of this omission is tested in \S\ref{check}.
Embedded in equation (\ref{masterinc}) are $N-1$ independent equations
for the $N-1$ distinct values of $\{I_j\}$.

We proceed to massage
equation (\ref{masterinc}) into a form that permits easy solution.
A dimensionless form of the $j^{\rm{th}}$ sub-equation reads

\begin{equation}
\frac{j-N}{2N-1} + \frac{4H(2N-1)M}{21\pi M_P \bar{i} J_2} \left(
\frac{\bar{a}}{R_p} \right)^2 \left( \frac{\bar{a}}{\Delta a} \right)^2
\sum_{k\neq j}^{2N-1} \frac{h_k}{k-j} q_{kj} = 0 \, ,
\label{master2}
\end{equation}

\ni where $\Delta a = a_{2N-1} - a_1$ is the total width of the ring
near quadrature, and $h_k = m_k / M$ is the fraction of the ring's
total mass contained in wire $k$.
Now each $q_{kj}$ is a linear combination of the $N-1$ independent
variables,

\begin{equation}
x_i \equiv q_{i,i+1} = \bar{a} \frac{I_i - I_{i+1}}{a_i - a_{i+1}}
\,\,\,\,\,\,\,\, \{ i = 1, \ldots, N-1 \} \, ,
\label{def}
\end{equation}

\ni as in

\begin{equation}
q_{j,j+l} = (1/l) \sum_{i=j}^{j+l-1} x_i \, .
\label{ident1}
\end{equation}

\ni Furthermore, for ring self-gravity to balance the
planetary quadrupole field, the inclination profile
must be reflection anti-symmetric about
the ring midline. Then

\begin{equation}
I_{N+j} - I_N = - (I_{N-j} - I_N) \,,
\label{ident2}
\end{equation}

\ni and if $j < N$ and $k > N$, then

\begin{equation}
q_{jk} = q_{jN} \frac{j-N}{j-k} + q_{2N-k,N} \frac{N-k}{j-k} \, .
\label{ident3}
\end{equation}

\ni It is straightforward to show that by virtue
of identities (\ref{ident1})--(\ref{ident3}), the summation
in equation (\ref{master2}) can be rewritten in terms of
the $\{x_i\}$ as

\begin{eqnarray}
\label{thebigone}
\sum_{k\neq j}^{2N-1} \frac{h_k}{k-j}q_{kj} & = & \sum_{i=1}^{i=j-1} x_i
\sum_{l=1}^{l=i} \frac{h_l}{(l-j)|l-j|} \\ \nonumber
 &  &  + \sum_{i=j}^{i=N-1} x_i \left[ \sum_{k=N+1}^{2N-1}
\frac{h_{2N-k}}{(k-j)^2} + \sum_{l=i+1}^{l=N} \frac{h_l}{(l-j)|l-j|} \right] \\
\nonumber
 &  &  + \sum_{i=1}^{i=N-1} x_i \sum_{l=1}^{l=i} \frac{h_l}{(2N-j-l)^2} \, .
\end{eqnarray}

\ni When equation (\ref{thebigone}) is substituted into equation
(\ref{master2}), we obtain a linear system of $N-1$ independent equations for
$N-1$
unknowns, $\{x_j\}$. These are solved numerically
using the LAPACK subroutines. The set of inclinations, $\{ I_j \}$, are
derived from $\{ x_j \}$ via (\ref{def}).

\placefigure{fig5}
\begin{figure}
\epsscale{0.8}
\vspace{-0.5in}
\plotone{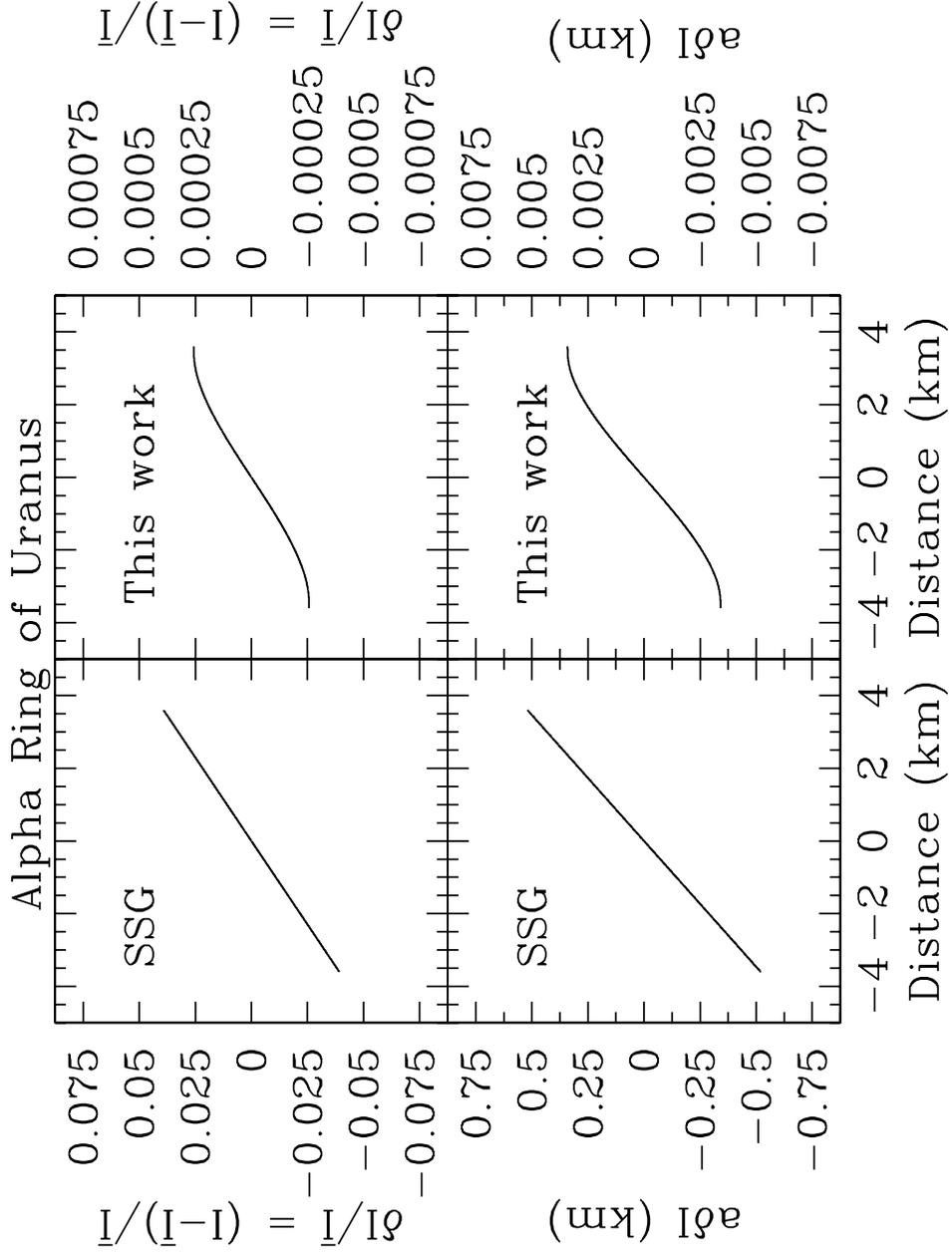}
\caption{Inclination profiles of the Uranian $\alpha$ ring using
two different mass models. Right-hand panels portray the inclination
profile using the surface density profile computed in \S\ref{surf},
which accounts for ring self-gravity, the planetary quadrupole
field, and interparticle collisions. Left-hand panels correspond
to inclination profiles computed under
the standard self-gravity model that does not
account for interparticle collisions. Upper panels describe
the fractional variation in inclination across the ring,
while lower panels plot the amplitude of the vertical warp across the ring.
The large masses predicted by our work yield much flatter
rings than those predicted by the standard self-gravity model;
we predict vertical warps of $\sim$7 meters across a ring
that is several kilometers wide.
\label{alphainc}}
\end{figure}

\placefigure{fig6}
\begin{figure}
\epsscale{0.8}
\vspace{-1in}
\plotone{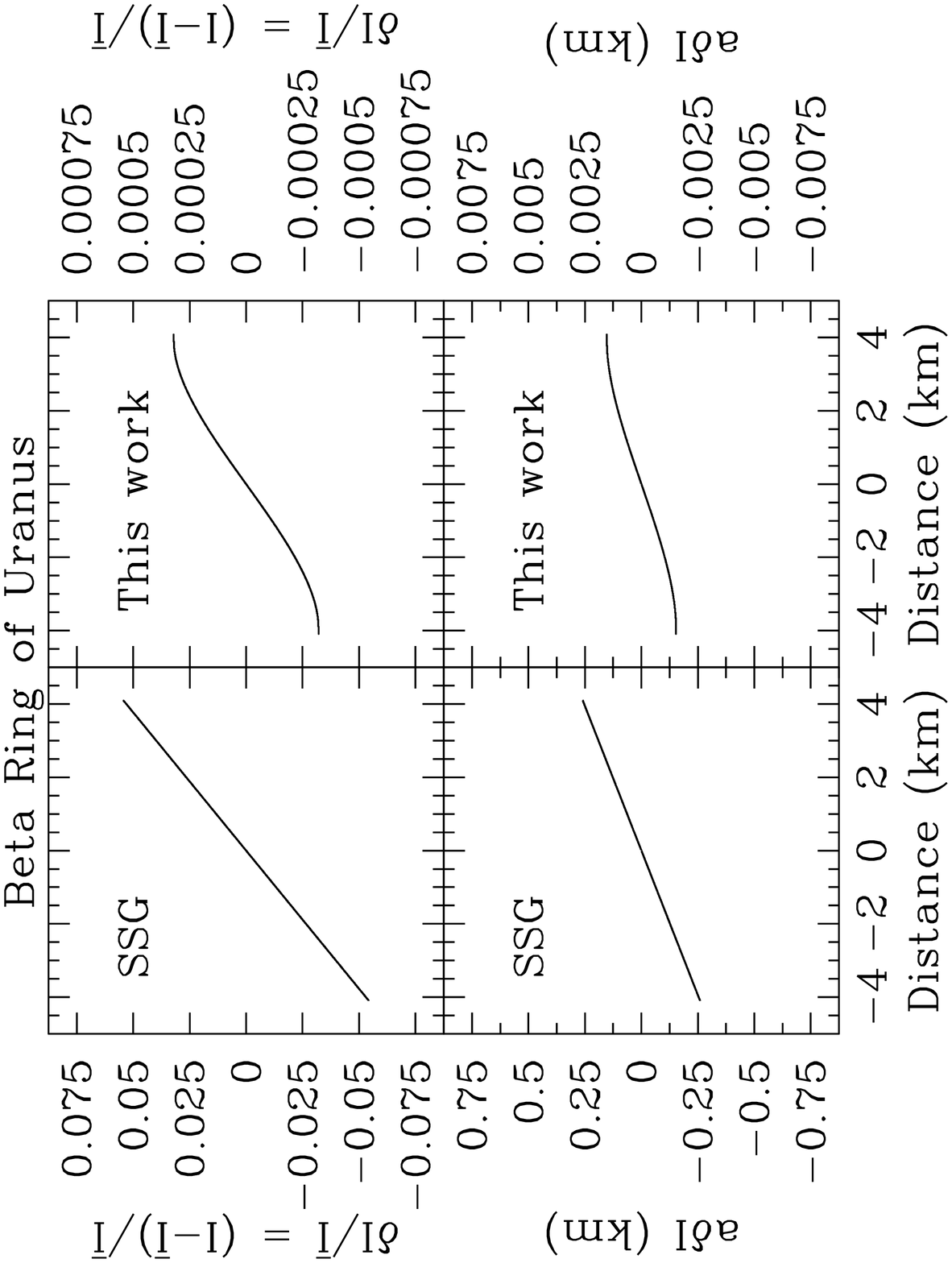}
\caption{Same as Figure \ref{alphainc}, but for the $\beta$ ring of Uranus.
Warp amplitudes of only $\sim$3 meters are predicted by our models.
\label{betainc}}
\end{figure}

Figures \ref{alphainc} and \ref{betainc} summarize the results
of our calculation for the $\alpha$ and $\beta$ rings of Uranus.
These are the only narrow rings whose mean inclinations, $\bar{I}$,
have been measured.
We predict inclination profiles that are much flatter than those predicted
by the standard self-gravity model. The larger ring masses
that arise from the inclusion of interparticle collisions
enable rings to maintain nodal alignment by warping their
geometry by relatively small amounts. The peak-to-peak amplitude
of the warp, defined as $a (I_{2N-1} - I_1)$, is
$\sim$7 meters for the $\alpha$ ring and $\sim$3 meters
for the $\beta$ ring. By contrast, the standard self-gravity
model, which employs wire masses derived without
regard to interparticle collisions, predicts warp amplitudes of order 1 km.

\placefigure{fig7}
\begin{figure}
\epsscale{0.8}
\vspace{-1in}
\plotone{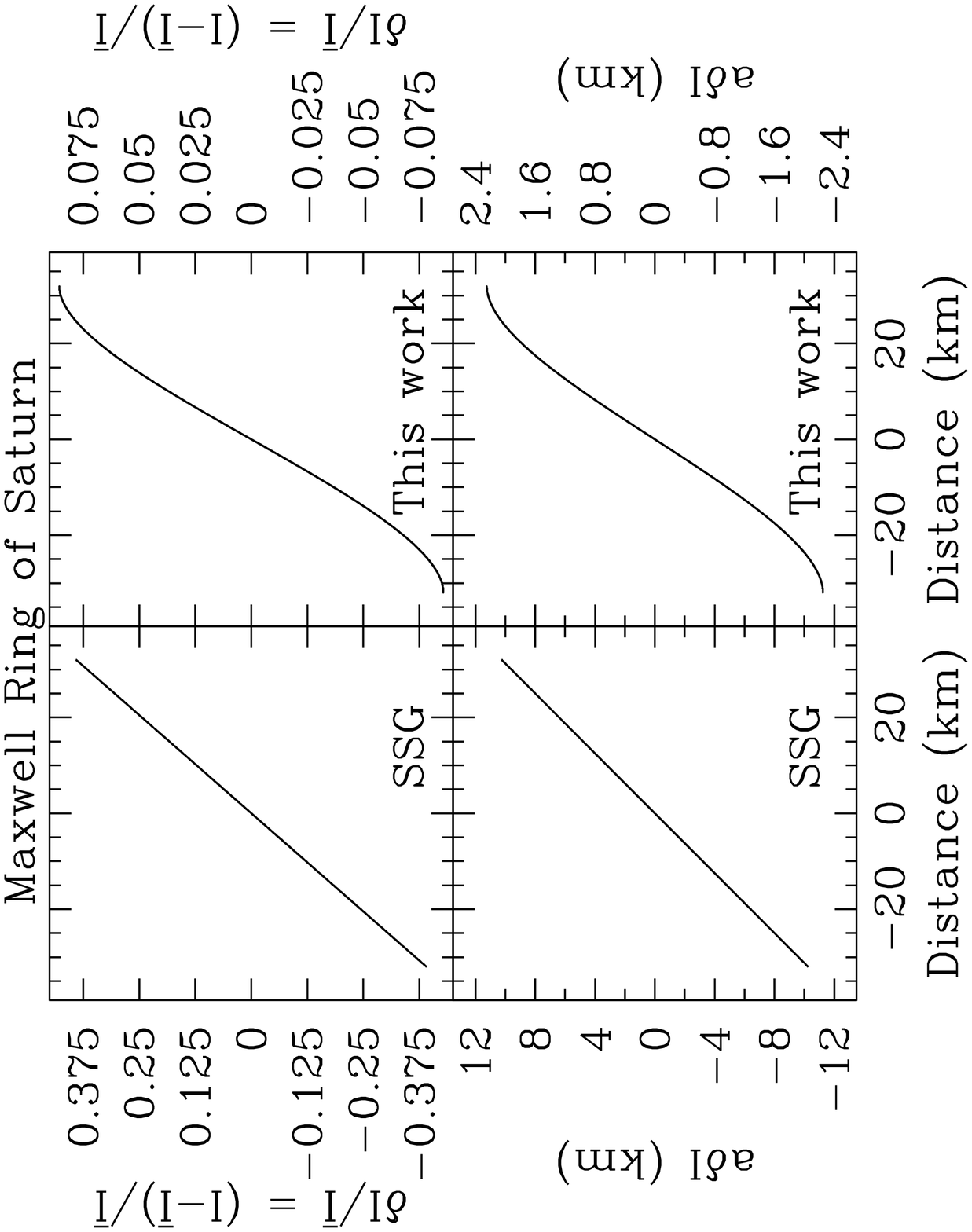}
\caption{Same as Figure \ref{alphainc}, but for the Maxwell ring of Saturn.
While the mean inclination, $\bar{I}$, is unknown for the
Maxwell ring, its value does not affect our calculation
of the fractional variation of inclination across the ring,
$(I-\bar{I})/\bar{I}$, shown in the upper panels. By contrast,
the absolute value of the vertical warp, $a \delta I = a (I-\bar{I})$,
scales linearly with $\bar{I}$; to compute the values 
displayed in the lower panels, we have assumed $\bar{I} = 3 \times 10^{-4}$
rad. Accounting for interparticle collisions raises the mass of the ring
and therefore reduces vertical warping; the effect is less
dramatic for the Maxwell ring than for the Uranian $\alpha$
ring because the radial width of the former is greater
than that of the latter, so that the solution is less sensitive
to boundary conditions.
\label{maxwellinc}}
\end{figure}

\placefigure{fig8}
\begin{figure}
\epsscale{0.8}
\vspace{-1in}
\plotone{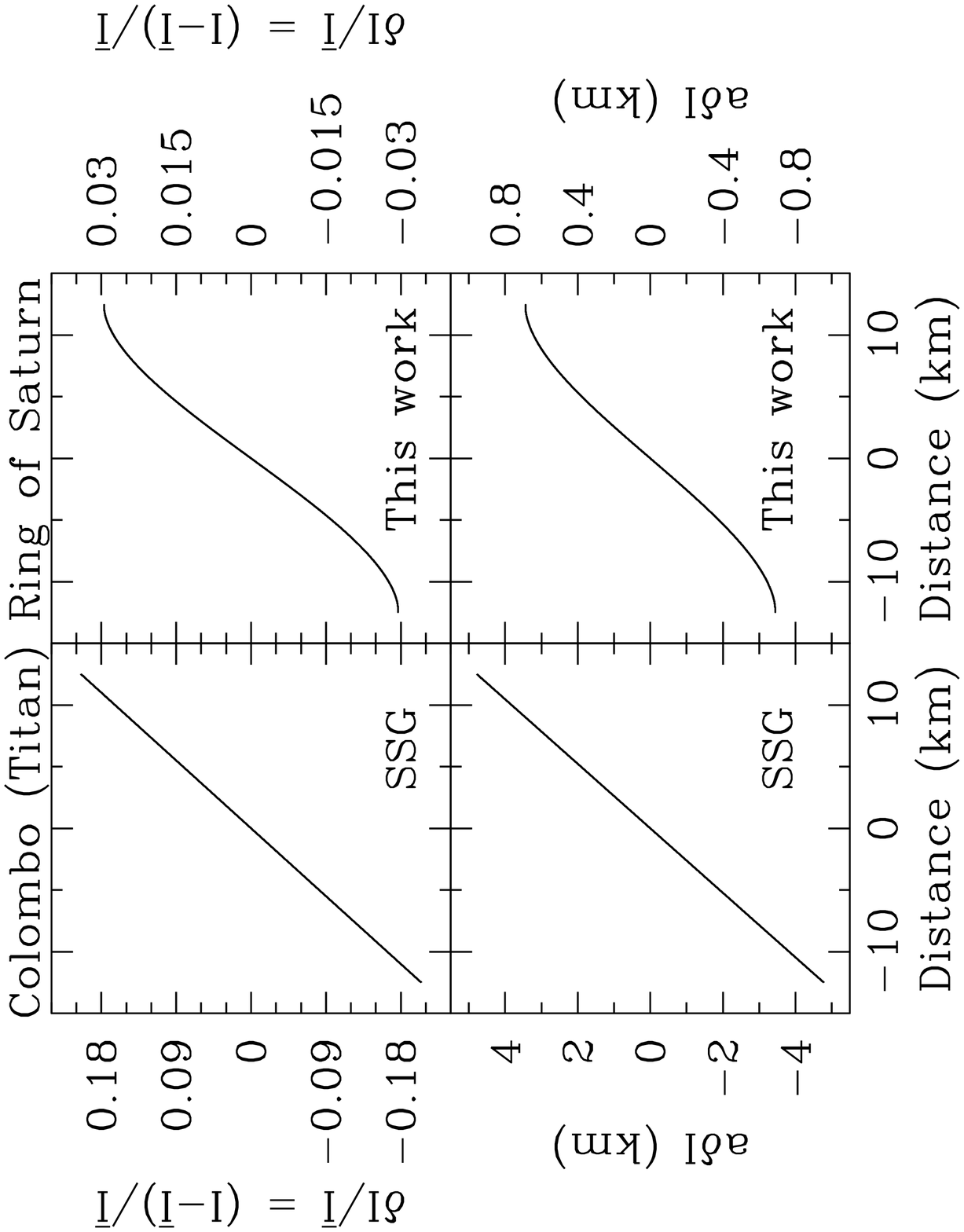}
\caption{Same as Figure \ref{maxwellinc}, but for the Colombo (Titan)
ring of Saturn.
In generating the bottom panels for $a\delta I = a (I-\bar{I})$,
we have assumed $\bar{I} = 3 \times 10^{-4}$ rad. The top panels
displaying the fractional variation of inclination
are independent of this assumed value.
\label{titaninc}}
\end{figure}

Figures \ref{maxwellinc} and \ref{titaninc} describe inclination
profiles for the Maxwell and Colombo ringlets of Saturn.
The mean inclinations, $\bar{I}$, of these rings with respect
to the local Laplacian plane have not been measured.
Fortunately, we have found by numerical experiment that
the {\it fractional} variation of inclination across the ring,
$(I - \bar{I})/\bar{I}$, is independent of the mean inclination.
The top panels of Figures \ref{maxwellinc} and \ref{titaninc}
should therefore be fairly accurate. Numerical experiment
also reveals that the {\it absolute} value of the warp
across the ring, $a (I_{2N-1} - I_1)$, scales linearly
with the unknown $\bar{I}$; in computing the absolute value
of the warp for the
bottom panels, we have simply guessed $\bar{I} = 3 \times 10^{-4}$ rad
for both rings, values comparable to their observed mean eccentricities.

Note that vertical warping is substantially greater for
the Saturnian Maxwell and Colombo rings than for the Uranian $\alpha$
and $\beta$ rings. Accounting for interparticle collisions
near ring edges always increases ring masses and flattens ring warps, but
such effects become less important as the ring width increases. The Maxwell
and
Titan rings are each more than 25 km across at quadrature,
while their Uranian counterparts are less than 7 km across.
Collisional stresses near ring edges are not as effectively communicated
across wider rings.

\subsection{Neglect of Collisions in Vertical Structure}
\label{check}

We check {\it a posteriori} the validity of our approximation in \S\ref{inca}
that interparticle collisions play no direct role in determining
the inclination profile. Collisions matter most near ring edges (CG00);
an upper limit for the collisional acceleration felt by a ring particle
lying with a mean free path, $\lambda \sim c_b/\bar{n}$,
of the ring edge is $c_b^2 / \lambda \sim c_b \, \bar{n}$.
The component of this acceleration that is perpendicular to the
orbit plane of this last ring particle is $c_b \, \bar{n} \, q_i$,
where $q_I = a \, (dI/da)$ is the inclination gradient evaluated
at the edge. This normal acceleration generates a rate of nodal
precession (relative to the collision-induced rate on the ring midline,
which is zero) of
$\sim$$c_b \bar{n} q_I / \bar{n} \bar{a} \bar{I} = c_b q_i / \bar{a} \bar{I}$.

We divide this maximum, collision-induced differential precession
rate by the differential rate of this last endwire
due to the quadrupole field of the planet.
For collisions to be unimportant, this dimensionless number should be
less than unity. For the Uranian $\alpha$ ring, it is

\begin{equation}
{\rm{Co}} \equiv \frac{8}{21} \frac{c_b q_I}{\bar{I}\bar{n}J_2\Delta a} \left(
\frac{\bar{a}}{R_p} \right)^2 = 0.2 
\left( \frac{c_b}{2 \cm \, \s^{-1}} \right)
\left( \frac{q_I}{2 \times 10^{-5}} \right) \, ,
\end{equation}

\ni where the value for $q_i$ at the ring edge is obtained from our computed
inclination profile. For the $\beta$ ring, we derive
$\rm{Co} \approx 0.2 (c_b / 2 \cm \, \s^{-1})$;
for both the Saturnian Maxwell and Colombo (Titan)
ringlets, $\rm{Co} \approx 0.1 (c_b / 2 \cm \, \s^{-1})$.
Note that these values of $\rm{Co}$ for the Saturnian ringlets
are independent of the assumed $\bar{I}$ since
$q_I / \bar{I} \propto \delta I / \bar{I}$ is independent of $\bar{I}$. 
We conclude based on these smallish values of $\rm{Co}$ that
collisions are of marginal direct importance to the inclination profiles of
our
ring models. As we recede from the ring
edge towards the ring interior,
collisions become even less important
as surface density gradients decline.


\section{STABILITY}
\label{stability}

Here we prove that circular, nodally locked ringlets
are linearly stable to perturbations to their
inclinations and nodes. A corresponding proof for the stability of
apsidally locked, co-planar rings would read similarly.
We account not only for ring self-gravity and the planetary quadrupole
field, but also for interparticle collisions, albeit in
a simplistic way.

We restrict ourselves to circular rings for simplicity and for this proof
only. Our hope is that accounting simultaneously
for ring eccentricity would not change our conclusions.
While this hope must be tempered by considerations
of a time-varying $\omega$ that we have not explored
in depth, we take some consolation from the Appendix in which we show
how a variable $\omega$ impacts force balance in the
vertical direction by seemingly modest amounts.

In \S \ref{onewire}, we derive the secular changes to the
inclination and node of a test particle due to the gravitational influence
of a nearby massive wire. In \S\ref{nwires}, we show how
narrow rings respond to small perturbations by oscillating
in a number of normal harmonic modes.

\subsection{One Wire and a Particle}
\label{onewire}

\begin{figure}
\epsscale{0.8}
\plotone{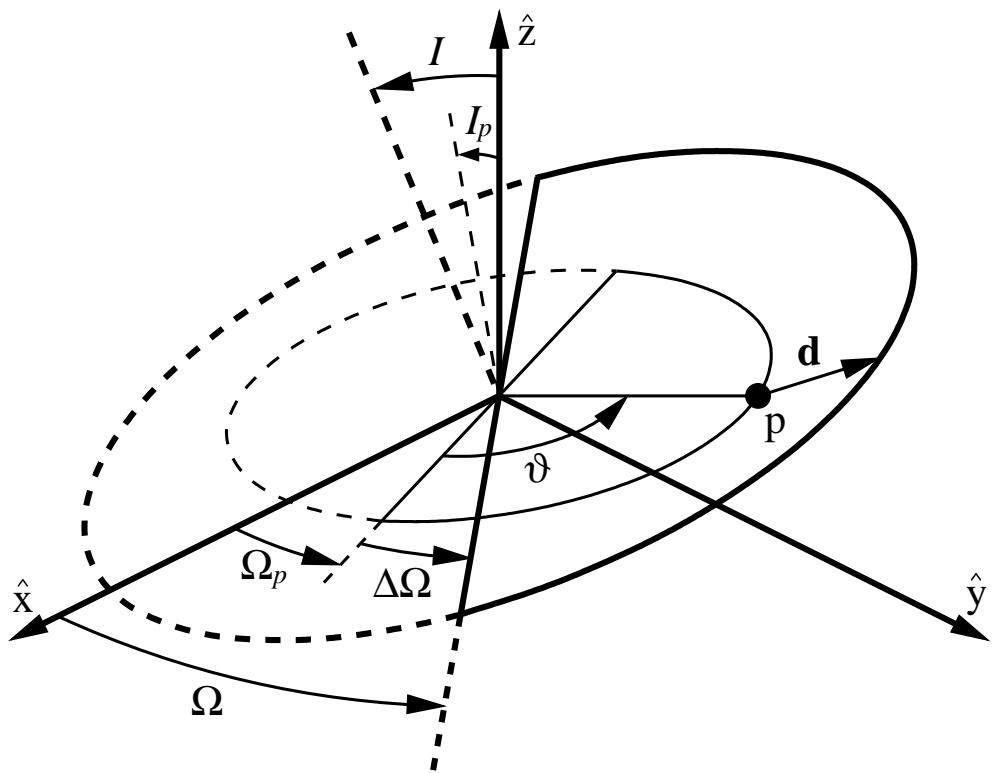}
\caption{The three-dimensional geometry of the problem. The central planet
rests at the origin, and the z-axis is aligned with the planet's spin axis. A
test particle
is located at point p and is displaced from its node on the planet's equator
plane by an angle $\vartheta$; $\mathbf{d}$ is
the perpendicular vector from the particle to the wire. Angles $\Omega$
and $I$ are
the longitude of ascending node and inclination of
the wire, respectively,
$\Omega_p$ and $I_p$ are the corresponding angles for the particle,
and $\Delta \Omega \equiv \Omega - \Omega_p$.
The inclinations, particle-wire separation, and misalignment of nodes
have all been exaggerated for clarity.
\label{c_loops}}
\end{figure}

Consider the perturbations induced by a massive, circular
wire on a nearby test particle. The equator plane of the planet
defines the reference plane, and $\mathbf{\hat{z}}$ is the unit vector
perpendicular to this plane.
Let $a_p$, $I_p$, and $\Omega_p$ be
the particle's semi-major axis, inclination, and longitude of ascending
node, while $a$, $I$, and $\Omega$ refer to the corresponding orbital
elements of the perturbing wire. Let $\vartheta$ denote the particle's
angular position away from the node, $\vartheta = \theta - \Omega_p$, where
$\theta$ is the
particle's longitude. Define $\Delta a \equiv a - a_p$,
$\Delta I \equiv I - I_p$, and $\Delta\Omega \equiv \Omega-\Omega_p$.
Figure \ref{c_loops} illustrates the geometry.

Assume that $|\Delta a| \ll a$,
$|\Delta I| \ll I$, and $|\Delta\Omega| \ll 1$.
If the wire has total mass $m$,
then its linear density is
\begin{equation} \label{c_rho}
	\rho = m / 2\pi a \, .
\end{equation}
Because the separation between the particle and the wire is small,
the wire can be approximated locally
as straight. The gravitational acceleration of the particle
induced by the wire is therefore
\begin{equation} \label{c_g}
	\mathbf{F} = 2 G \rho \mathbf{d} / d^2,
\end{equation}
where $\mathbf{d}$ is the perpendicular vector from the particle to
the wire, $d = |\mathbf{d}|$, and $G$ is the gravitational constant.

\placefigure{fig9}

\placefigure{fig10}
\begin{figure}
\epsscale{1.0}
\plotone{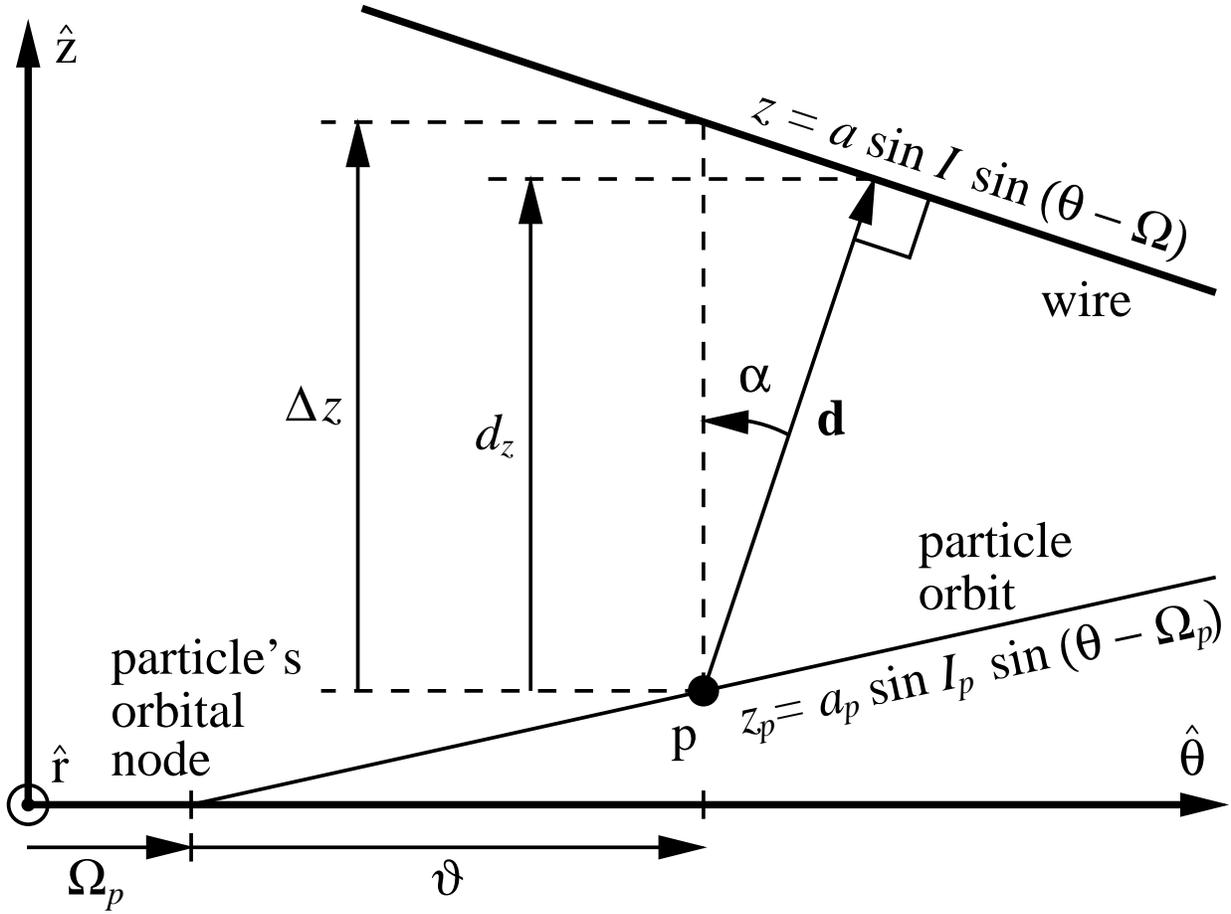}
\caption{A two-dimensional projection of Figure \ref{c_loops}
that plots height above the planet's equator plane
versus longitude on that plane.
To achieve an exaggerated difference
in orbits, we make $\Delta \Omega$ large and negative for this figure.
Because the angle $\alpha$ is $\mathcal{O}(I)$,
$d_z = \Delta z \left[1 - \mathcal O (I^2)\right]$.
\label{c_proj}}
\end{figure}

As depicted in Figure \ref{c_proj}, the vertical distance from the particle
to the wire at fixed longitude is
\begin{eqnarray}
	\Delta z  & = & a \sin I \sin(\vartheta - \Delta\Omega) - a_{p} \sin
I_{p}\sin\vartheta \nonumber \\
	& = & \Delta (aI) \sin \vartheta - aI\Delta\Omega \cos \vartheta,
	\label{c_deltaz}
\end{eqnarray}
where $\Delta(aI) = aI - a_p I_p$. Borderies et al.\ (1983b) assume
$\Delta(aI) \approx a\Delta I$, which is equivalent to
$\Delta a/a \ll \Delta I/I$; since we have derived inclination profiles
in \S\ref{inca} for which their assumption is not valid,
we keep our more accurate expression.
Referring again to Figure \ref{c_proj}, we note that since
$\mathcal{O}(\alpha) = \mathcal{O}(I)$,

\begin{equation}
d_z \equiv \mathbf{d} \cdot \mathbf{\hat{z}} = \Delta z [ 1 - \mathcal{O}(I^2)
] \, .
\end{equation}

\ni Furthermore,

\begin{equation}
\label{simplegeo}
d^2 = d_z^2 + \Delta a^2 \, .
\end{equation}

The changes in $I_p$ and $\Omega_p$ due to $\mathbf{F}$ are given to leading
order by (see, e.g., Murray \& Dermott 1999)
\begin{mathletters}\begin{eqnarray}
	\label{c_dodt} \frac{d\Omega_{p}}{dt}(\vartheta) & = & \frac{F_z \sin
\vartheta}{n_{p} a_{p} I_{p}} \\
	\label{c_didt} \frac{dI_p}{dt}(\vartheta) & = & \frac{F_z \cos
\vartheta}{n_{p} a_{p}},
\end{eqnarray}
where\end{mathletters} $n_p$ is the particle's mean motion,
$F_z \equiv \mathbf{F} \cdot \mathbf{\hat{z}} = 2G\rho d_z/d^2 \simeq 2G\rho
\Delta z / d^2$ is the vertical component of the perturbing force,
and we have neglected terms of order $I^2$. Substituting eqs.\
(\ref{c_rho}), (\ref{c_g}), (\ref{c_deltaz}), and (\ref{simplegeo})
into (\ref{c_dodt}) and (\ref{c_didt}) yields
\begin{mathletters}\begin{eqnarray}
\label{c_dodt2}\frac{d\Omega_{p}}{dt}(\vartheta)  & = & C
	\frac{\frac{\Delta (aI)}{aI} \sin^2 \vartheta
	- \Delta \Omega \sin \vartheta \cos \vartheta}
	{1 + \left(\frac{\Delta (aI)}{\Delta a} \sin \vartheta
	- \frac{aI\Delta\Omega}{\Delta a} \cos \vartheta \right)^2}, \\
\label{c_didt2}\frac{dI_p}{dt}(\vartheta)  & = & C
	\frac{\frac{\Delta (aI)}{a} \sin \vartheta \cos \vartheta - I \Delta \Omega
\cos^2 \vartheta}
	{1 + \left(\frac{\Delta (aI)}{\Delta a} \sin \vartheta
	- \frac{aI\Delta\Omega}{\Delta a} \cos \vartheta \right)^2},
\end{eqnarray}
where\end{mathletters}
\begin{equation} \label{c_defc}
	C = \frac{m}{\pi M_P} {\left(\frac{a}{\Delta a}\right)}^2 n
\end{equation}
and $n$ is the mean motion at semi-major axis $a$. Where appropriate,
we have neglected differences between $a_p$ and $a$.

To obtain the secular changes in $I_p$ and $\Omega_p$, we average
$d\Omega_{p}/dt$ and $dI_p/dt$
over one full
orbit of the test particle; to wit,
$X = (1/2\pi)\int_0^{2\pi}X(\vartheta) \, d\vartheta$. To
simplify this integral, first define $Q$ and $\gamma$ such that
\begin{mathletters}\begin{eqnarray}
\label{c_defqc}Q \cos \gamma & = & \frac{\Delta (aI)}{\Delta a} \\
\label{c_defqs}Q \sin \gamma & = & \frac{a I \Delta \Omega}{\Delta a}.
\end{eqnarray}
Then\end{mathletters} eqs.\ (\ref{c_dodt2}) and (\ref{c_didt2}) can be
re-written
\begin{mathletters}\begin{eqnarray}
\label{c_dodt3}\frac{d\Omega_p}{dt}(\vartheta) & = & C
	\left( \frac{\Delta a}{a I} \right) \frac{Q \sin \vartheta \sin (\vartheta -
\gamma)}
	{1 + Q^2 \sin^2 (\vartheta - \gamma)} \\
\label{c_didt3}\frac{dI_p}{dt}(\vartheta) & = & C
	\left( \frac{\Delta a}{a} \right) \frac{Q \cos \vartheta \sin (\vartheta -
\gamma)}
	{1 + Q^2 \sin^2 (\vartheta - \gamma)} \, .
\end{eqnarray}
Averaging\end{mathletters} these equations over $\vartheta$ yields
\begin{mathletters}\begin{eqnarray}
\label{c_avgo}\left(\frac{d\Omega_p}{dt}\right)_{G} & = &
	\phantom- C \left( \frac{\Delta a}{a I} \right) H(-Q^2) \,Q \cos \gamma
	 = C H(-Q^2)\left(\frac{\Delta I}{I} + \frac{\Delta a}{a}\right) \\
\label{c_avgi}\left(\frac{dI_p}{dt}\right)_{G} & = &
	-C \left( \frac{\Delta a}{a} \right) H(-Q^2) \,Q \sin \gamma
	= -C H(-Q^2)I\Delta\Omega \, ,
\end{eqnarray}
where\end{mathletters} the subscript $G$ denotes wire gravity and
\begin{equation}
\label{c_defh}H(-Q^2) = \frac{\sqrt{Q^2 + 1}-1}{Q^2\sqrt{Q^2+1}}.
\end{equation}
The forms of $C$ and $H$ are chosen to be consistent with eq.\ (6) of
Borderies et al.\ (1983b).

\subsection{N Self-Gravitating Wires with Oblate Planet and Collisions}
\label{nwires}
Now consider the interaction between $N$ circular wires.
We demonstrate that
the equilibrium configuration of the wires is linearly stable.
For this purpose, we do not take the wires to be equally spaced in
semi-major axis, but rather define their spacing such that all wires have
equal mass; the wires must therefore be more closely packed in
regions of higher surface density. This de-composition by mass
rather than by semi-major axis is critical to our proof
of stability.
Denote the total ring mass by $M$, so that each wire has mass $M/N$.
Let wire $j$ have
semi-major axis $a_j$, longitude of ascending node $\Omega_j$, and inclination
$I_j$.
Define mean elements by $\bar{a} \equiv \sum a_j/N$, $\bar{\Omega} \equiv \sum
\Omega_j /N$,  and
$\bar{I} \equiv \sum I_j/N$; let $\bar{n}$ denote the Keplerian mean motion at
semi-major axis $\bar{a}$.
As before, we assume that variations in
orbital elements across the ring are small: for any wire $j$,
$|a_j - \bar{a}| \ll \bar{a}$, $|I_j - \bar{I}| \ll \bar{I}$, and $|\Omega_j -
\bar{\Omega}| \ll 1$.
We adapt eqs.\ (\ref{c_avgo})
and (\ref{c_avgi}) to write down the time rates of change of $\Omega_j$
and $I_j$ due to the gravitational
attraction of wire $k$:
\begin{mathletters}\begin{eqnarray}
\label{c_ndo1}\left(\frac{d\Omega_j}{dt}\right)_{G(k)} & = & \phantom-
C_{jk}H(-Q_{jk}^2)
		\left(\frac{I_k-I_j}{\bar{I}}+\frac{a_k-a_j}{\bar{a}}\right)\\
\label{c_ndi1}\left(\frac{dI_j}{dt}\right)_{G(k)} & = &
-C_{jk}H(-Q_{jk}^2)I(\Omega_k-\Omega_j) \, ,
\end{eqnarray}
where\end{mathletters}
\begin{equation}
	C_{jk} = C_{kj} = \frac{M}{\pi M_PN}\left(\frac{\bar{a}}{a_k - a_j}\right)^2
\!\bar{n}
\end{equation}
\begin{equation}
	Q_{jk}^2 = Q_{kj}^2 = \left(\frac{a_kI_k-a_jI_j}{a_k-a_j}\right)^2 +
	\left(\frac{\bar{a} \bar{I} (\Omega_k-\Omega_j)}{a_k-a_j}\right)^2.
\end{equation}

Interparticle collisions are assimilated as follows.
According to equation (\ref{collacc}), 
the collisional repulsion between two neighboring wires is inversely
proportional to
the distance between them, and the resulting acceleration is directed
oppositely to that of inter-wire gravity. This prescription implies that
collisional forces act effectively as anti-gravitational
forces, having the same dependence on inter-wire separation
as self-gravity but with an opposite sign.
Thus, we can account for
collisions by prepending a factor $g_{jk}$ to the right-hand sides
of eqs.\ (\ref{c_ndo1}) and
(\ref{c_ndi1}). The factor $g_{jk}$ represents the effects of collisions;
the particular value
of $g_{jk}$ depends on the local velocity dispersion; it is always
$< 1$,
and it is negative where collisional repulsion is stronger than gravitational
attraction.

Finally, consider the central planet's quadrupole field, which causes no
secular change in
$\{I_j\}$ but induces precession of the nodes $\{\Omega_j\}$. The effect is
described to leading order by
\begin{equation}\label{c_prec}
	\left(\frac{d\Omega_j}{dt}\right)_Q = - \frac{3}{2} J_2 R_p^2 \sqrt{GM_P}
a_j^{-7/2} \, .
\end{equation}
A linear expansion yields
\begin{equation}
	\left(\frac{d\Omega_j}{dt}\right)_Q = \frac{d\bar{\Omega}}{dt} +
		\frac{21}{4}J_2\left(\frac{R_p}{\bar{a}}\right)^2\! \bar{n}\, \frac{a_j -
\bar{a}}{\bar{a}},
\end{equation}
where $d\bar{\Omega}/dt=-\frac{3}{2} J_2 R_p^2 \sqrt{GM_P} \bar{a}^{-7/2}$.

The total rates of change of $\Omega_j$ and $I_j$ due to all other wires
read as
\begin{mathletters}\begin{eqnarray}
	\frac{d\Omega_j}{dt} & = & \sum_{k \neq j} g_{jk}C_{jk}H(-Q_{jk}^2)
		\left(\frac{I_k-I_j}{\bar{I}}+\frac{a_k-a_j}{\bar{a}}\right) \nonumber \\
\label{c_ndo2}& & + \frac{d\bar{\Omega}}{dt} +
\frac{21}{4}J_2\left(\frac{R_p}{\bar{a}}\right)^2 \!\bar{n}\, \frac{a_j -
\bar{a}}{\bar{a}}\\
\label{c_ndi2} \frac{dI_j}{dt} & = & -\sum_{k \neq
j}g_{jk}C_{jk}H(-Q_{jk}^2)I(\Omega_k-\Omega_j).
\end{eqnarray}
\end{mathletters}

\ni Eq.\ (\ref{c_ndi2}) implies that $\bar{I}$ is conserved.
Now denote the equilibrium inclination of wire $j$ by $I_j^0$, and define
\begin{equation}
	\Phi_j = \frac{I_j - I_j^0}{\bar{I}} \, .
\end{equation}

\ni Equations (\ref{c_ndo2}) and (\ref{c_ndi2}) are expressed more concisely in
terms of
$\Phi_j$:
\begin{mathletters}\begin{eqnarray}
\label{c_ndo3}\frac{d\Omega_j}{dt} & = & \phantom-\sum_{k \neq
j}g_{jk}C_{jk}H(-Q_{jk}^2)
		(\Phi_k - \Phi_j) + \frac{d\bar{\Omega}}{dt}\\
\label{c_ndi3}\frac{d\Phi_j}{dt} & = & -\sum_{k \neq
j}g_{jk}C_{jk}H(-Q_{jk}^2)(\Omega_k-\Omega_j).
\end{eqnarray}\end{mathletters}

\ni The terms in eq.\ (\ref{c_ndo2}) proportional to $(a_k-a_j)/\bar{a}$ and
$(a_j-\bar{a})/\bar{a}$ have been absorbed into $I_j^0$ (i.e., $\Phi_j$)
and $I_k^0$ (i.e., $\Phi_k$).

As a final simplification, define $N \times 1$ column vectors $\vec x$, $\vec
y$, and $\vec u$
with elements $x_j = \Omega_j$, $y_j = \Phi_j$, and $u_j = 1$. Equations
(\ref{c_ndo3}) and (\ref{c_ndi3}) can then be written in matrix form:
\begin{mathletters}\begin{eqnarray}
\label{c_mdx}\frac{d}{dt}\, \vec x & = & \phantom- K \vec y + \frac{d
\bar{\Omega}}{dt}\vec u, \\
\label{c_mdy}\frac{d}{dt}\, \vec y & = & -K \vec x,
\end{eqnarray}
where\end{mathletters} $K$ is an $N\times N$ symmetric matrix whose
off-diagonal
entries are given by
\begin{equation}
	K_{jk} = K_{kj} = g_{jk}C_{jk}H(-Q_{jk}^2),
\end{equation}
and whose diagonal entries equal
\begin{equation}
	K_{jj} = - \sum_{k \neq j}g_{jk}C_{jk}H(-Q_{jk}^2).
\end{equation}
Note that $K$ is symmetric only because we have chosen our wires to have equal
mass.
Furthermore, we can take $H$ to be constant because the fractional variation
in this quantity is of order $Q^2$ times the fractional variation in $Q$,
and $Q \sim q_I \ll 1$ from our equilbrium model.
Finally, we assume that variations in $g_{jk}$ are small.
Then $K$ is a constant matrix and eqs.\ (\ref{c_mdx}) and (\ref{c_mdy}) are
linear.

Since $K$ is symmetric, it possesses $N$ orthogonal eigenvectors $\vec v_0,
\ldots, \vec v_{N-1}$ with
real associated eigenvalues $\chi_0, \ldots, \chi_{N-1}$.
Inspection reveals that $\vec v_0 = \vec u$ and $\chi_0 = 0$.
Thus the complete solution to eqs.\ (\ref{c_mdx})
and (\ref{c_mdy}) is
\begin{mathletters}\begin{eqnarray}
\label{c_mx}\vec x & = & \sum_{n=1}^{N-1} c_n \vec v_n \sin \left[ \chi_n (t -
\phi_n) \right]
		+ \frac{d\bar{\Omega}}{dt}(t-t_0)\vec u, \\
\label{c_my}\vec y & = & \sum_{n=1}^{N-1} c_n \vec v_n \cos \left[ \chi_n (t -
\phi_n) \right],
\end{eqnarray}
where\end{mathletters} the $2N-2$ constants $c_1, \ldots, c_{N-1}$ and $\phi_1,
\ldots, \phi_{N-1}$
are determined from the initial relative inclinations and longitudes of
ascending nodes.
The two remaining constants are $t_0$ and $\bar{I}$, where $t_0$ is set by the
mean longitude
of ascending node. Equations (\ref{c_mx}) and (\ref{c_my}) imply that an
inclined, circular
ring responds to small perturbations to the inclinations and nodes of its
constituent
ringlets in the presence of self-gravity, planetary oblateness, and (a
simplified
prescription of) interparticle collisions by oscillating in a number of normal
modes.

These modes resemble the vibrations of a string with free boundary
conditions and variable density.
A preliminary numerical study in which we modelled the ring
with $g_{jk} = 1$ suggests that the
fundamental (lowest, non-zero) frequency
is

\begin{equation}\label{c_l2}
	\chi_1 \sim \frac{2M}{\pi M_P} {\left(\frac{\bar{a}}{\Delta a}\right)}^2
\bar{n} \,,
\end{equation}

\ni where $\Delta a$ is the full width of the ring. For this mode,
half the ring precesses nodally in one direction relative to the ring midline,
while the other half precesses in the other direction.
For the Maxwell ringlet, $2\pi / \chi_1 \sim 8$ yr.
This same study indicates that $\chi_j = 2 \chi_1 j$
for $j > 10$. We will use this numerical model in our discussion
below regarding the possibility of exciting a high order mode
by using force variations that arise from a
time-varying $\omega$ (see also the Appendix).


\section{SUMMARY AND DISCUSSION}
\label{discussion}

Our crude and prescriptive treatment of streamline dynamics
near resonantly perturbed ring boundaries
prevents us from claiming great accuracy in our computed
surface density and inclination profiles. Nonetheless,
we believe the following conclusions to be robust:

\begin{enumerate}

\item Narrow rings have total masses of order $10^{19}\g$.

\item Their surface densities increase towards their edges.

\item The Uranian $\alpha$ and $\beta$
rings are vertically warped by a height of order 10 m.
The radially wider Saturnian Maxwell and Titan rings are warped
by a height of order 1 ($\bar{I} / 3 \times 10^{-4} \rad$) km.
An equivalent and more model-independent statement is that the fractional
variation of inclination is order $10^{-3}$
across the Uranian ringlets and of order
$10^{-1}$ across their Saturnian analogues.

\item Nodally (apsidally) locked rings are linearly stable
to perturbations to their inclinations (eccentricities)
and nodes (apses).

\end{enumerate}

\ni Analysis of stellar occultation data by the Cassini spacecraft
may test predictions 1--3 for the Maxwell and Colombo ringlets
of Saturn.

For a ring to maintain apsidal and nodal alignment, its argument
of pericenter must precess. This means that where the ring
is ``pinched''---in other words, its pericenter, the narrowest
portion of the ring, where self-gravitational and collisional forces are
strongest---circulates relative to the node of the ring
on the equator plane of the planet.
Such circulation must introduce time-variable forces
into the vertical equations of motion for the ring;
these time-variable forces have a frequency of
$2\dot{\pomega}$, or about
0.08 cycles per day for the Saturn Maxwell ringlet.
Might these time-variable forces drive bending waves
across the ring?
First note that the driving frequency
is more than 2 orders of magnitude greater
the fundamental bending frequency, $\chi_1$.
Therefore any mode that is excited
will have a wavelength considerably shorter than the overall width
of the ring. A preliminary study suggests that
the radial wavelength of the bending mode whose
natural frequency matches the driving frequency
is of order one kilometer for the Saturnian ringlets,
a length scale possibly within reach of the Cassini camera.
We defer a more thorough investigation of this possibility,
including estimates of the amplitude of the mode, to future study.
In the event that Cassini finds the vertical structure
of a narrow ring to vary over a period of a few weeks
and over kilometer lengthscales,
one might look to the periodic stresses associated with
a time-varying $\omega$ for the cause.
Ultimately, measurements and a theoretical understanding
of any modes excited might
constrain the mass of the ring independently of optical depth soundings.

We have solved for the inclination gradient of a given ring
by first finding the surface density profile. A key input
for this procedure was the eccentricity profile of the ring, $e(a)$,
which we obtained from the (spatially averaged) observations.
While this procedure is physically self-consistent, it
begs the question of where the eccentricity profile
originated. Another way of saying this is to note
that our equilibrium solutions are not the only ones possible
in principle. For example, the steps of our procedure
could be reversed: we could begin by positing an inclination
gradient, solve next for the masses required to maintain nodal
alignment, and conclude by deriving the eccentricity gradient
required to maintain apsidal alignment. We did not adopt this
procedure because the observations supplied the eccentricity
gradient and not the inclination gradient. Why narrow rings
self-organize themselves to exhibit
dimensionless eccentricity gradients of order $q_e = \bar{a} de/da \sim 0.5$
and dimensionless inclination gradients that are
significantly smaller---of order
$q_I = \bar{a} dI/da \sim 10^{-1}$--$10^{-3}$
for the Saturnian cases and of order
$10^{-3}$--$10^{-5}$ for the Uranian cases---is unclear.

Many issues remain unresolved in our understanding of
narrow rings. In the more than twenty years since their
discovery, we still ask the following questions:

\begin{enumerate}
\item Are narrow rings primordial? According to classical
theories of ring shepherding, attendant satellites extend
the viscous spreading time of a ring by a factor of order the mass
ratio between the shepherd and the ring. This extension
factor is modest for the Uranian $\epsilon$ ring, whose
mass is thought to be comparable to that of its shepherds (CG00).
The situation is likely to be similar for other
rings, especially given the large ring masses that we are deriving.
Estimates for the spreading time of the $\epsilon$ ring range
from $3 \times 10^4$ yr to $2 \times 10^8$ yr,
all of which are shorter than the
age of the solar system (Chiang 2003). But if narrow rings are not
formed from primordial circumplanetary disks, how can we
explain the tiny mean inclinations exhibited by the Uranian
rings with respect to the equator plane of their severely
oblique host planet? New estimates of ring lifetimes
based on non-classical theories of shepherding (see, e.g.,
Goldreich \& Porco 1987, and Borderies, Goldreich, \& Tremaine 1984)
would be welcome.
\item Why do all narrow rings exhibit positive and not negative
eccentricity gradients?
\item Why is the inclination gradient smaller in magnitude
than the eccentricity gradient?
\end{enumerate}

We might hope to answer these questions by modelling more
carefully the distortion of streamlines near ring edges
by shepherd satellites.

\acknowledgements
It is a pleasure to thank Joanne Cohn for introducing C.J.C. to E.I.C.
We thank also Carolyn Porco for faxing pages from her thesis to us,
and an anonymous referee for a helpful report that prompted
us to examine our statements regarding bending modes more carefully.
This work was supported in part by a Space Grant Summer Fellowship awarded
to C.J.C. by the Space Sciences Laboratory at Berkeley. E.I.C.
acknowledges support by
National Science Foundation Planetary Astronomy Grant 
AST-0205892 and Hubble Space Telescope Theory Grant HST-AR-09514.01-A.

\appendix
\section{Time-varying $\omega$}
Here we explore the effect of a time-varying argument
of pericenter, $\omega$, on our equation of motion (\ref{igrav}) for
nodal precession in the presence of ring self-gravity.
Consider two elliptical wires orbiting a spherical planet,
one placed in an arbitrary reference plane,
and a second placed at an inclination, $\Delta I$,
relative to that plane. Take the longitudes of pericenter
of the two wires to be equal, $\pomega_1 = \pomega_2 = 0$.
The longitude of ascending node of the second wire is $\Omega$.
How sensitive is the orbit-averaged rate of nodal precession of the second
wire, $\langle \dot{\Omega} \rangle$, to $\Omega = -\omega$?

At every true anomaly, $f$, of the second wire, the perpendicular
distance between the second wire and the first is

\begin{equation}
\label{append}
d \approx \sqrt{ (\Delta a)^2 ( 1 - q_e \cos f )^2 + a^2 (\Delta I)^2 \sin^2
(f-\Omega) } \, ,
\end{equation}

\ni where we have assumed that the difference in semi-major axes
of the wires, $\Delta a$, is much smaller than the semi-major axis
of either wire, $a$. The eccentricity gradient between the wires
is $q_e \equiv a \Delta e / \Delta a$, where $\Delta e$ is the difference
in wire eccentricities. The first term under the square root
in equation (\ref{append}) represents the in-plane separation between
the wires, while the second term represents the out-of-plane contribution.
Re-write (\ref{append}) as

\begin{equation}
\label{append1}
d \approx \Delta a \sqrt{ (1-q_e \cos f)^2 + q_I^2 \sin^2 (f-\Omega) } \, ,
\end{equation}

\ni where $q_I \equiv a \Delta I / \Delta a$.

By Gauss's equation, the instantaneous
rate of nodal precession of a test particle
whose orbit coincides with the second wire reads

\begin{equation}
\label{append2}
\dot{\Omega} = \frac{F_z \sin (f-\Omega)}{na\Delta I} \, ,
\end{equation}

\ni where $n$ equals the mean motion of the second wire, $F_z$ equals
the vertical acceleration felt by the test particle,

\begin{equation}
\label{append3}
F_z = -\frac{2G\rho a \Delta I \sin (f-\Omega)}{d^2} \, ,
\end{equation}

\ni and $\rho$ equals the linear mass density of the first wire of mass $m$,

\begin{equation}
\label{append4}
\rho = \frac{m}{2\pi a} \, .
\end{equation}

\ni We have dropped corrections to (\ref{append4}) due to wire
eccentricity because the acceleration (\ref{append3}) is already
small in $\Delta I$. Time-averaging $\dot{\Omega}$
over one orbital period yields

\begin{equation}
\label{append5}
\langle \dot{\Omega} \rangle = -\frac{m}{\pi M_P} n \frac{a}{\Delta a}
\frac{q_I}{\Delta I} B(q_e, q_I, \Omega) \, ,
\end{equation}

\begin{equation}
\label{append6}
B = \frac{1}{2\pi} \int_0^{2\pi} \frac{\sin^2 (f-\Omega)}{(1-q_e \cos f)^2 +
q_I^2 \sin^2(f-\Omega)} df \, .
\end{equation}

\ni Equation (\ref{append5}) is identical in form to equation (\ref{igrav}).
The sensitivity of $\langle \dot{\Omega} \rangle$ to $\omega = - \Omega$ is
contained in the integral, $B$. Numerical integration of (\ref{append6})
reveals that $B \in [0.613, 0.896]$ for $q_e = 0.5$ and $q_I = 0.1$.
If $q_e = 0.5$ and $q_I = 0$, then $B \in [0.619, 0.920]$.
The variation of force with varying $\omega$ is restricted to less
than a factor of 1.5. In \S\ref{equilibrium}, we fixed $B = 0.77$
for a maximum 24\% fractional error; the error so accrued
is less than the error introduced by our prescriptions for
force balance near ring boundaries. However, the force variation is large
enough that it may excite observable
short-wavelength bending waves across the ring,
as we discuss in \S\ref{discussion}.


\newpage
\begin{deluxetable}{cccccccccc}
\tablewidth{0pc}
\tabletypesize{\scriptsize}
\tablecaption{Observed and Derived Parameters of Narrow Eccentric
Rings\label{param}}
\tablehead{
\colhead{Ring}  &  \colhead{Planet} &
\colhead{$\bar{a}$(km)\tablenotemark{a}} & \colhead{$\Delta
a$(km)\tablenotemark{a}} &
\colhead{$\bar{e}\, (\times 10^3)$\tablenotemark{a}}   &
\colhead{$(\bar{q}_e)$\tablenotemark{a}} &
\colhead{$\bar{I}\, (\times 10^3)$\tablenotemark{a}}   &
\colhead{$(\bar{q}_I)$\tablenotemark{b}} &
\colhead{$(\Delta I/\bar{I})$\tablenotemark{b}} &
\colhead{$M$($10^{19}$g)\tablenotemark{b}}
}

\startdata

$\alpha$ & Uranus & 44718 & 7.15 & 0.761 & 0.472 & 0.265 & 0.00095 & 0.00057 &
0.467  \\

$\beta$  & Uranus & 45661 & 8.15 & 0.442 & 0.370 & 0.089 & 0.00037 & 0.00075 &
0.495 \\

Maxwell  & Saturn & 87491 & 64 & 0.34 & 0.46 & (0.3) & (0.070) & 0.17 & 5.00
\\

Colombo (Titan) & Saturn & 77871 & 25 & 0.26 & 0.44 & (0.3) & (0.055) & 0.059 &
1.64 \\

\tablenotetext{a}{Uranian ring values taken from Tables I and VII of French et
al. (1991). Saturnian ring values taken from Table IIId of Porco (1983) and
Table 1 of Porco (1990). Values for $\bar{I}$ in parentheses are guessed.}
\tablenotetext{b}{Derived from this work. Values in parentheses
for $\bar{q}_I \equiv \bar{a} \Delta I / \Delta a$,
where $\Delta I$ is the difference in orbital
inclinations between inner and outer ring edges, scale
linearly with the guessed value of $\bar{I}$.}
\enddata
\end{deluxetable}

\newpage
\begin{deluxetable}{cccccc}
\tablewidth{0pc}
\tablecaption{Theoretical Model Parameters\label{param2}}
\tablehead{
\colhead{Ring}  &
\colhead{$c_i\,(\rm{cm} \, \s^{-1})$} & \colhead{$c_b\,(\rm{cm} \, \s^{-1})$}
&
\colhead{$w_r$(km)\tablenotemark{a}} & \colhead{$\lambda$(km)\tablenotemark{b}}
& \colhead{$N$}
}

\startdata

$\alpha$ & 0.1 & 2.0 & 0.44 & 0.079 & 1000  \\

$\beta$  & 0.1 & 2.0 & 0.45 & 0.081 & 1000  \\

Maxwell  & 0.1 & 3.0 & 0.56 & 0.126 & 3000  \\

Colombo (Titan) & 0.1 & 2.0 & 0.50 & 0.071 & 3000 \\

\tablenotetext{a}{Resonant width, computed as
$w_r = \bar{a} \sqrt{M_{sat}/M_P}$,
where the mass of an individual (as yet unobserved) shepherd satellite equals
$M_{sat} = 8.4 \times 10^{18} \gm$ for the Uranian ringlets
and $M_{sat} = 2.3 \times 10^{19} \gm$ for the Saturnian ringlets.}
\tablenotetext{b}{Mean free path near ring edge, computed as
$\lambda = c_b / \bar{n}$, where $\bar{n}$ is the average
mean motion of the ring.}
\enddata
\end{deluxetable}

\end{document}